\documentclass[aps,showpacs,twocolumn]{revtex4}
\usepackage{epsfig}
\usepackage{color}
\usepackage{amsmath}
\usepackage{amssymb}
\usepackage{bbm}

\newcommand{\be}{\begin{equation}}
\newcommand{\ee}{\end{equation}}
\newcommand{\bea}{\begin{eqnarray}}
\newcommand{\beaa}{\begin{eqnarray*}}
\newcommand{\eea}{\end{eqnarray}}
\newcommand{\eeaa}{\end{eqnarray*}}

\numberwithin{equation}{section}

\begin{document}

\title{\bf\Large {Towards the theory of ferrimagnetism}}

\author{Naoum Karchev\cite{byline}}

\affiliation{Department of Physics, University of Sofia, 1164 Sofia, Bulgaria}

\begin{abstract}
Two-sublattice ferrimagnet, with spin-$s_1$ operators $\bf{S_{1i}}$ at the sublattice $A$ site and spin-$s_2$ operators $\bf{S_{2i}}$ at the sublattice $B$ site, is considered.
The magnon of the system, the transversal fluctuation of the total magnetization, is a complicate mixture of the transversal fluctuations of the sublattice $A$ and $B$ spins. As a result, the magnons' fluctuations suppress in a different way the magnetic orders of the $A$ and $B$ sublattices and one obtains two phases. At low temperature $(0,T^*)$ the magnetic orders of the $A$ and $B$ spins  contribute to the
magnetization of the system, while at the high temperature  $(T^*,T_N)$, the magnetic order of the spins with a weaker intra-sublattice exchange is suppressed by magnon fluctuations, and only the spins with stronger intra-sublattice exchange has non-zero spontaneous magnetization.
The $T^*$ transition is a transition between two spin-ordered phases in contrast to the transition from spin-ordered state to disordered state ($T_N$-transition). There is no additional symmetry breaking, and the Goldstone boson has a ferromagnetic dispersion in both phases.
A modified spin-wave theory is developed to describe the two phases. All known Neel's anomalous $M(T)$ curves are reproduced, in particular that with "compensation point". The theoretical curves are compared with experimental ones for sulpho-spinel $MnCr2S_{4-x}Se_{x}$  and rare earth iron garnets.

\end{abstract}

\pacs{75.50.Bb, 75.30.Ds, 75.60.Ej, 75.50.-y}

\maketitle

\section {\bf Introduction}

The notions of ferrimagnetism and ferrimagnetic materials were introduced by L. Neel \cite{Neel} for materials which spontaneous magnetization is a resultant of two or more components of non-parallel magnetic moments. Using a molecular field theory L. Neel predicted the nature of magnetization $M$ versus temperature $T$. Anomalous $M(T)$ curves arise due to the fact that  each of the magnetic moments approaches its own saturation value as a different function of temperature. In general there is nothing to limit the number of components. The simplest model consists of two alternating sublattices of unequal and antiparallel moments, with three molecular field coefficients utilized to describe the exchange field effects: one ferromagnetic coefficient for each sublattice and a third for antiferromagnetic interaction. In many cases, the experimental curves are used to determine the molecular field coefficients. The most striking feature of the anomalous $M(T)$ curves is the possibility "of compensation point", a temperature $T_c$ at which the
magnetic moments of two sublattices are equal and opposite, so that the magnetization of the system is equal to zero $M(T_c)=0$. The phenomenon of ferrimagnetism has been subject of extensive experimental investigation since its discovery. The phenomenological Neel's standpoint has been confirmed by many authors (see the review articles \cite{Wolf,Buschow,HBMM2,HBMM3,Belov}).

The earliest application of spin waves to ferrimagnet was made by H. Kaplan \cite{swKaplan} to calculate the dispersion $\omega_k$ for the spinel-type ferrite. For small wave vector $\textbf{k}$ he found a quadratic relation $\omega_k=D\textbf{k}^2$, where $D$ is a constant, as for ferromagnetism. To simplify the calculations Kaplan neglected the intra-sublattice exchange compared with inter-sublattice one, so that the calculations do not correspond to any real ferrimagnet.

In the present paper I consider two-sublattice ferrimagnet, with spin-$s_1$ operators $\bf{S_{1i}}$ at the sublattice $A$ site and spin-$s_2$ operators $\bf{S_{2i}}$ at the sublattice $B$ site.
The Hamiltonian of the system is
\bea \label{ferri1}
 H & = & - J_1\sum\limits_{\ll ij \gg _A } {{\bf S}_{1i}
\cdot {\bf S}_{1j}}\,-\,J_2\sum\limits_{\ll ij \gg _B } {{\bf S}_{2i}
\cdot {\bf S}_{2j}}\nonumber \\
& & +\,J \sum\limits_{\langle ij \rangle} {{\bf S}_{1i}
\cdot {\bf S}_{2j}}
  \eea where the sums are over all sites of a three-dimensional cubic lattice:
$\langle i,j\rangle$ denotes the sum over the nearest neighbors, $\ll i,j \gg _A$ denotes the sum over the sites of A sublattice, $\ll i,j \gg _B$ denotes the sum over the sites of B sublattice.
The first two terms  describe the ferromagnetic Heisenberg intra-sublattice
exchange $J_1>0, J_2>0$, while the third term describes the inter-sublattice exchange which is antiferromagnetic $J>0$.

The true magnons of a two-spin system are transversal fluctuations of the total magnetization which includes both the magnetization of sublattice $A$ spins and magnetization of sublattice $B$ spins.
The magnon excitation is a complicate mixture of the transversal
fluctuations of the sublattice $A$ and $B$ spins. As a result the magnons' fluctuations suppress
in a different way the magnetic orders on the different sublattices and one obtains two phases.
At low temperature $(0,T^*)$ the magnetic orders of the $A$ and $B$ spins  contribute to the
magnetization of the system, while at the high temperature  $(T^*,T_N)$ the magnetic order of the spins with a weaker intra-sublattice exchange is suppressed by magnon fluctuations and only the spins with stronger intra-sublattice exchange has non-zero spontaneous magnetization.
At first sight the result that there is a temperature interval where a magnetic order is formed only by spins on one of the sublattices seems to be counterintuitive because the
moment on the one of the sublattices build an effective magnetic field, which due to inter-sublattice exchange
interaction leads to finite magnetization on the other sublattice.
This is true in the classical limit. In the quantum case the spin
wave fluctuations suppress the $A$ and $B$ magnetic orders at different temperatures $T^*$ and $T_N$ as a
result of different interaction of magnons with  sublattices' spins.
The $T^*$ transition is a transition between two spin-ordered phases in contrast to the transition from spin-ordered state to disordered state ($T_N$-transition). There is no additional symmetry breaking and the Goldstone boson
has a ferromagnetic dispersion in both phases.

A modified spin-wave theory is developed to describe the two phases. By means of this method of calculation the thermal variation of magnetization is calculated for three cases: $(s_1>s_2)$\,\,$J_1\gg J_2$,\,\, $(s_1>s_2)$\,\,$J_2\gg J_1$ and $(s_1=s_2)$\,\,$J_1\gg J_2$. All known Neel's anomalous $M(T)$ curves are reproduced including the curve with a famous "compensation point" (second case).

An important issue is the experimental detection of $T^*$ transition. In the case when $T^*$ is the temperature at which itinerant electrons start to form magnetic order the transition
demonstrates itself through the change in the $T$ dependence of resistivity.
It is well known that the onset of magnetism in itinerant systems is
accompanied with strong anomaly in resistivity \cite{itinerant}. One expects the same
phenomena when the itinerant electrons in ferrimagnets form magnetic order. There are experimental results
which support the above interpretation of the $T^*$ transition.

The paper is organized as follows: In Sec. II, a spin wave theory of the model Eq.(\ref{ferri1}) is presented. It is shown the non-adequacy of the customary spin-wave theory for description of the high temperature phase. Sec. III is devoted to the development of a modified spin wave theory. The calculations are accomplished for different choices of the model's parameters. The theoretical $M(T)$ curves are compared with experimental ones for sulpho-spinel $MnCr2S_{4-x}Se_{x}$ \cite{HBMM3} and rare earth iron garnets \cite{Wolf} and satisfying coincidence is obtained.  A summary in Sec. IV concludes the paper.

\section{\bf Spin-wave theory}

To study a theory with Hamiltonian Eq.(\ref{ferri1}) it is convenient to introduce Holstein-Primakoff representation for the spin
operators
\bea\label{ferri2} & &
S_{1j}^+ = S^1_{1j} + i S^2_{1j}=\sqrt {2s_1-a^+_ja_j}\,\,\,\,a_j \nonumber \\
& & S_{1j}^- = S^1_{1j} - i S^2_{1j}=a^+_j\,\,\sqrt {2s-a^+_ja_j}
\\ & & S^3_{1j} = s_1 - a^+_ja_j \nonumber \eea
when the sites $j$ are from sublattice $A$, and
\bea\label{ferri3} & &
S_{2j}^+ = S^1_{2j} + i S^2_{2j}=-b^+_j\,\,\sqrt {2s_2-b^+_jb_j}\nonumber \\
& & S_{2j}^- = S^1_{2j} - i S^2_{2j}=-\sqrt {2s_2-b^+_jb_j}\,\,\,\,b_j
\\ & & S^3_{2j} = -s_2 + b^+_jb_j \nonumber \eea
when the sites $j$ are from sublattice $B$. The operators $a^+_j,\,a_j$ and  $b^+_j,\,b_j$ satisfy the Bose commutation relations. In terms of
the Bose operators and keeping only the quadratic terms the effective
Hamiltonian Eq.(\ref{ferri1}) adopts the form
\bea\label{ferri4}
 H & = & s_1 J_1\sum\limits_{\ll ij \gg _A }\left( a^+_i a_i\,+\,a^+_j a_j\,-\,a^+_j a_i\,-\,a^+_i a_j\right) \nonumber \\
 & + & s_2 J_2\sum\limits_{\ll ij \gg _B }\left( b^+_i b_i\,+\,b^+_j b_j\,-\,b^+_j b_i\,-\,b^+_i b_j\right) \\
 & + &  \sum\limits_{\langle ij \rangle}\left[ s_1 J b^+_j b_j+s_2 J a^+_i a_i-
 J \sqrt{s_1s_2}\left( a^+_i b^+_j+a_i b_j \right)\right] \nonumber
 \eea

 To proceed one rewrites the Hamiltonian in momentum space representation
 \be \label{ferri5}
 H = \sum\limits_{k\in B_r}\left [\varepsilon^a_k\,a_k^+a_k\,+\,\varepsilon^b_k\,b_k^+b_k\,-
 \,\gamma_k \left (a_k^+b_k+b_k^+a_k \right )\,\right ],
\ee
where the wave vector $k$ runs over the reduced  first Brillouin zone $B_r$ of a
cubic lattice. The dispersions are given by equalities
\bea
\label{ferri6} & & \varepsilon^a_k\,=\,4s_1\,J_1\,\varepsilon_k
\,+\,6s_2\,J
\nonumber \\
\\
& & \varepsilon^b_k\,=\,4s_2\,J_2\,\varepsilon_k \,+\,6s_1\,J \nonumber \eea
with
\bea\label{ferri7}
& & \varepsilon_k = 6-\cos(k_x+k_y)-\cos(k_x-k_y) - \cos (k_x+k_z) \nonumber \\
& - & \cos(k_x-k_z)
 - \cos(k_y+k_z) - \cos(k_y-k_z) \eea
and
\be \label{ferri8}
\gamma_k\,=\,2J\,\sqrt{s_1\,s_2}\,\left(\cos k_x +\cos k_y + \cos k_z \right) \ee

To diagonalize the Hamiltonian one introduces new Bose fields
$\alpha_k,\,\alpha_k^+,\,\beta_k,\,\beta_k^+$ by means of the
transformation
\bea \label{ferri9} & &
a_k\,=u_k\,\alpha_k\,+\,v_k\,\beta^+_k\qquad
a_k^+\,=u_k\,\alpha_k^+\,+\,v_k\,\beta_k
\nonumber \\
\\
& & b_k\,=\,u_k\,\beta_k\,+\,v_k\,\alpha^+_k\qquad
b_k^+\,=\,u_k\,\beta_k^+\,+\,v_k\,\alpha_k
\nonumber \eea where the coefficients of the transformation $u_k$ and $v_k$ are real function of the wave vector $k$
\bea \label{ferri11} &
&u_k\,=\,\sqrt{\frac 12\,\left (\frac
{\varepsilon^a_k+\varepsilon^b_k}{\sqrt{(\varepsilon^a_k+\varepsilon^b_k)^2-4\gamma^2_k}}\,+\,1\right
)}\nonumber \\
\\
& & v_k\,=\,sign (\gamma_k)\,\sqrt{\frac 12\,\left (\frac
{\varepsilon^a_k+\varepsilon^b_k}{\sqrt{(\varepsilon^a_k+\varepsilon^b_k)^2-4\gamma^2_k}}\,-\,1\right
)}\nonumber \eea
The transformed Hamiltonian adopts the form \be
\label{ferri12} H = \sum\limits_{k}\left
(E^{\alpha}_k\,\alpha_k^+\alpha_k\,+\,E^{\beta}_k\,\beta_k^+\beta_k\,+\,E^0_k\right),
\ee
with new dispersions \bea  \label{ferri13} & & E^{\alpha}_k\,=\,\frac
12\,\left [
\sqrt{(\varepsilon^a_k\,+\,\varepsilon^b_k)^2\,-\,4\gamma^2_k}\,-\,\varepsilon^b_k\,+\,\varepsilon^a_k\right] \nonumber \\
\\
& & E^{\beta}_k\,=\,\frac
12\,\left [
\sqrt{(\varepsilon^a_k\,+\,\varepsilon^b_k)^2\,-\,4\gamma^2_k}\,+\,\varepsilon^b_k\,-\,\varepsilon^a_k\right]
\nonumber \eea and vacuum energy
\be\label{ferri14}
 E^{0}_k\,=\,\frac
12\,\left [
\sqrt{(\varepsilon^a_k\,+\,\varepsilon^b_k)^2\,-\,4\gamma^2_k}\,-\,\varepsilon^b_k\,-\,\varepsilon^a_k\right]\ee

For all values of $k$
$\sqrt{(\varepsilon^a_k\,+\,\varepsilon^b_k)^2\,-\,4\gamma^2_k}\geq|\varepsilon^b_k\,-\,\varepsilon^a_k|$,
and the dispersions are nonnegative $ E^{\alpha}_k\geq 0,\, E^{\beta}_k \geq 0$.

For definiteness I choose $s_1>s_2$.
With these parameters, $\beta_k$-boson is a gapped excitation with gap
\be\label{ferri17a} E^{\beta}_0 = 6 J (s_1-s_2), \ee while $\alpha_k$-boson is the long-range \textbf{(magnon)} excitation in the two-spin system. Near the zero wave vector
\be\label{ferri17b}E^{\alpha}_k\approx \rho k^2 \ee
where the
spin-stiffness constant is
\be\label{ferri17c} \rho\,=\,\frac {8 s_1^2 J_1\,+\,8 s_2^2 J_2\,+\,2 s_1 s_2 J}{s_1\,-\,s_2}\ee
The spontaneous magnetization of the system $M$
is a sum of the spontaneous magnetization on the two sublattices $M\,=\,M^A\,+\,M^B$, where
\bea \label{ferri17d}
M^A & = & <S^3_{1j}> \,\,\, j\,\, is\,\, from\,\, sublattice\,\, A \nonumber \\
\\
M^B & = & <S^3_{2j}> \,\,\, j\,\, is\,\, from\,\, sublattice\,\, B \nonumber \eea
In terms of the Holstein-Primakoff bosons ($a_k$ and $b_k$) the magnetization  adopts the form
\bea \label{ferri18}& & M^A\,=\,s_1\,-\,\frac 1N
\sum\limits_{k\in B_r}<a^+_k a_k>
\nonumber \\
\\
& & M^B\,=\,-s_2\,+\,\frac 1N \sum\limits_{k\in B_r}<b^+_k
b_k> \nonumber \eea
where $N_A=N_B=N$ are the numbers of sites on sublattices A and B.
Finally one can rewrite
$M^A$ and $M^B$ in terms of the $\alpha_k$ ($\alpha_k^+$) and $\beta_k$ ($\beta_k^+$) excitations
\bea \label{ferri19}& & M^A\,=\,s_1\,-\frac 1N
\sum\limits_{k\in B_r}\left [u^2_k\,<\alpha^+_k \alpha_k>\,+\,v^2_k\,<\beta_k \beta^+_k>\right]\nonumber
\\
\\
& & M^B\,=\,-s_2\,+\frac 1N \sum\limits_{k\in B_r}\left
[u^2_k\,<\beta^+_k \beta_k>\,+\,v^2_k\,<\alpha_k \alpha^+_k> \right]\nonumber \eea
Then, the
magnetization of the system adopts the form \be \label{ferri20}
M\,=\,s_1\,-\,s_2-\frac 1N \sum\limits_{k\in B_r}\left [<\alpha^+_k \alpha_k>\,-\, <\beta^+_k \beta_k> \right]\ee
In equations (\ref{ferri19},\ref{ferri20})
\bea\label{ferri21}
& & <\alpha^+_k \alpha_k>=\frac {1}{e^{E^{\alpha}_k/T}-1},\quad <\alpha_k \alpha^+_k>=1+<\alpha^+_k \alpha_k>
 \nonumber \\
\\
& & <\beta^+_k \beta_k>=\frac {1}{e^{E^{\beta}_k/T}-1},\quad <\beta_k \beta^+_k>=1+<\beta^+_k \beta_k>
\nonumber \eea

We consider a theory with Hamiltonian Eq. (\ref{ferri1}), where the sums are over all sites of a three-dimensional cubic lattice with space size $a=1$. The two equivalent sublattices A and B are face centered cubic (fcc) lattices with space size $2a=2$. The coordinates \,$k_x,k_y,k_z$\, of the wave vector $k$ are coordinates in the basis $\hat{x},\hat{y},\hat{z}$, which are primitive vectors of the cubic lattice. To implement the integrations over the wave vector it is more convenient to switch to coordinates \,$q_1,q_2,q_3$\, in the basis $\hat{A},\hat{B},\hat{C}$, which are primitive vectors of the reciprocal lattice to the fcc lattice with space size equal to 2.
\bea\label{ferri23}
\hat{A} & = & \pi \left(\hat{x}\,+\,\hat{y}\,-\,\hat{z}\right) \nonumber \\
\hat{B} & = & \pi \left(-\hat{x}\,+\,\hat{y}\,+\,\hat{z}\right)  \\
\hat{C} & = & \pi \left(\hat{x}\,-\,\hat{y}\,+\,\hat{z}\right) \nonumber\eea
To that end we utilize the transformation
\bea\label{ferri24}
k_x & = & \pi q_1\,-\,\pi q_2\,+\,\pi q_3 \nonumber \\
k_y & = & \pi q_1\,+\,\pi q_2\,-\,\pi q_3  \\
k_z & = & -\pi q_1\,+\,\pi q_2\,+\,\pi q_3 \nonumber \eea
The Jacobian of the transformation is equal to $4\pi^3$, so that $d^3k/(2\pi)^3= d^3q/2$.
When the wave vector $k$ runs the reduced Brillouin zone $k\in B_r$, the new vector runs the cube $\quad 0 \leq q_l \leq 1$.
Finally we introduce the more convenient coordinates $p_1\,=\,\pi q_1;\quad p_2\,=\,\pi q_;\quad p_3\,=\,\pi q_3$.
They run the interval $[0,\pi]$ and $d^3k/(2\pi)^3= d^3p/2\pi^3$. In terms of the new coordinates the functions of the wave vector (\ref{ferri7}) and (\ref{ferri8}) adopts the form
\bea\label{ferri25}
\varepsilon_p & = & 6-\cos(2p_1)-\cos(2p_2)-\cos (2p_3) \nonumber \\
& - & \cos2(p_1-p_2)-\cos2(p_1-p_3)-\cos2(p_2-p_3) \nonumber \\
\gamma_p & = & 2J\,\sqrt{s_1\,s_2}\,\left[\cos(-p_1+p_2+p_3) +\cos(p_1-p_2+p_3)\nonumber \right. \\ & + & \left.\cos(p_1+p_2-p_3) \right] \eea

The magnon excitation-$\alpha_k$ in the effective theory
Eq.(\ref{ferri12}) is a complicate mixture of the transversal
fluctuations of the $A$ and $B$ spins. As a result the magnons' fluctuations suppress
in a different way the magnetic order on sublattices $A$ and $B$.
Quantitatively this depends on the coefficients $u_k$ and
$v_k$ in Eqs.(\ref{ferri19}). The magnetization depends on the
dimensionless temperature $T/J$ and dimensionless parameters
$s_1,\,s_2,\,J_1/J$ and $J_2/J$. For parameters $s_1=1.5,\,s_2=1,\,J_1/J=0.47$ and $J_2/J=0.005$ the
functions $M(T/J)$, $M^A(T/J)$ and $M^B(T/J)$ are depicted in
Fig.1.  The upper (green) line is the sublattice $A$ magnetization, the bottom (red) line is the sublattice $B$ magnetization
and the middle (blue) line is the magnetization of the system. The figure
shows that the magnetic order on sublattice $B$ (red line) is
suppressed first, at temperature $T^{*}/J=20$. Once suppressed,
the magnetic order can not be restored at temperatures above $T^{*}$
because of the increasing effect of magnon fluctuations. Hence, the
sublattice $B$ magnetization should be zero above
$T^{*}$. It is evident from Fig.1, that this is not the result within
customary spin-wave theory.
\begin{figure}[!ht]
\epsfxsize=9cm 
\epsfbox{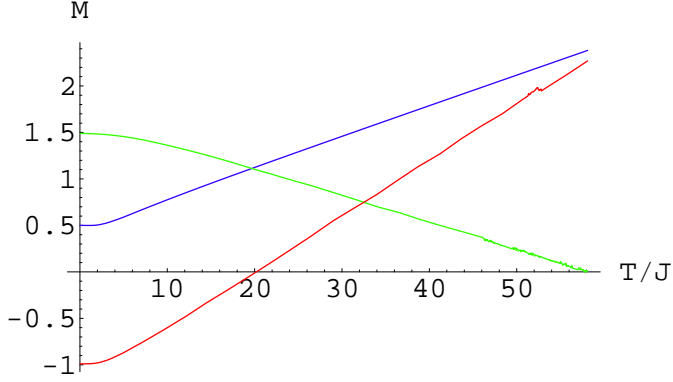} \caption{(color online)\,Temperature dependence of the ferromagnetic
moments: $M$ (blue line)-the magnetization of the system, $M^A$
(green line)-sublattice $A$ magnetization, $M^B$ (red
line)-sublattice $B$ magnetization for parameters
$s_1=1.5,\,s_2=1,\,J_1/J=0.47$ and $J_2/J=0.005$:\,\textbf{ spin-wave theory}}\label{fig1}
\end{figure}

\section{\bf Modified spin-wave theory}

To solve the problem we use the idea on description of paramagnetic
phase of 2D ferromagnets ($T>0$) by means of modified spin-wave
theory \cite{Takahashi1,Takahashi2}. In the simplest version the spin-wave
theory is modified by introducing a new parameter which enforce the
magnetization of the system to be equal to zero in paramagnetic
phase.

We consider two-sublattice system and to enforce the magnetic
moments on the two sublattices to be equal to zero in paramagnetic pase
we introduce two parameters $\lambda_1$ and $\lambda_2$. The new Hamiltonian is obtained from the old one adding two new terms
to the Hamiltonian Eq.(\ref{ferri1}) \be
\label{ferri26} \hat{H}\,=\,H\,-\,\sum\limits_{i\in A}
\lambda_1 S^3_{1i}\,+\,\sum\limits_{i\in B} \lambda_2 S^3_{2i} \ee
In momentum space
the new Hamiltonian adopts the form \be \label{ferri27}
 \hat{H} = \sum\limits_{k\in B_r}\left [\hat{\varepsilon}^a_k\,a_k^+a_k\,+\,\hat{\varepsilon}^b_k\,b_k^+b_k\,-
 \,\gamma_k\,(b_k a_k+b_k^+ a_k^+)\right] \ee
where the new dispersions are
 \be \label{ferri28}
\hat{\varepsilon}^a_k\,=\varepsilon^a_k\,+\,\lambda_1, \qquad
\hat{\varepsilon}^b_k\,=\varepsilon^b_k\,+\,\lambda_2.\ee
Utilizing the same
transformation Eq.(\ref{ferri9}) with parameters
\bea \label{ferri29} &
&\hat{u}_k\,=\,\sqrt{\frac 12\,\left (\frac
{\hat{\varepsilon}^a_k+\hat{\varepsilon}^b_k}{\sqrt{(\hat{\varepsilon}^a_k+\hat{\varepsilon}^b_k)^2-4\gamma^2_k}}\,+\,1\right
)}\nonumber \\
\\
& &\hat{v}_k\,=\,sign (\gamma_k)\,\sqrt{\frac 12\,\left (\frac
{\hat{\varepsilon}^a_k+\hat{\varepsilon}^b_k}{\sqrt{(\hat{\varepsilon}^a_k+\hat{\varepsilon}^b_k)^2-4\gamma^2_k}}\,-\,1\right
)}\nonumber \eea
one obtains the Hamiltonian in diagonal forma
\be
\label{ferri30} \hat{H} = \sum\limits_{k\in B_r}\left
(\hat{E}^{\alpha}_k\,\alpha_k^+\alpha_k\,+\,\hat{E}^{\beta}_k\,\beta_k^+\beta_k+\hat{E}^0_k\right),
\ee where
\bea\label{ferri31} & & \hat{E}^{\alpha}_k\,=\,\frac
12\,\left [
\sqrt{(\hat{\varepsilon}^a_k\,+\,\hat{\varepsilon}^b_k)^2\,-\,4\gamma^2_k}\,-\,\hat{\varepsilon}^b_k\,+\,\hat{\varepsilon}^a_k\right] \nonumber \\
& & \hat{E}^{\beta}_k\,=\,\frac
12\,\left [
\sqrt{(\hat{\varepsilon}^a_k\,+\,\hat{\varepsilon}^b_k)^2\,-\,4\gamma^2_k}\,+\,\hat{\varepsilon}^b_k\,-\,\hat{\varepsilon}^a_k\right]\\
& & \hat{E}^{0}_k\,=\,\frac
12\,\left [
\sqrt{(\hat{\varepsilon}^a_k\,+\,\hat{\varepsilon}^b_k)^2\,-\,4\gamma^2_k}\,-\,\hat{\varepsilon}^b_k\,-\,\hat{\varepsilon}^a_k\right]\nonumber\eea

We have to do some assumptions for the parameters $\lambda_1$ and
$\lambda_2$ to ensure correct definition of the Bose theory.
For that purpose it is convenient to represent the parameters
$\lambda_1$ and $\lambda_2$ in the form \be \label{ferri32}
\lambda_1\,=\,6 J s_2 \mu_1\,-\,6 J s_2,\quad
\lambda_2\,=\,6 J s_1 \mu_2\,-\,6 J s_1. \ee In terms of the new
parameters $\mu_1$ and $\mu_2$ the dispersions
$\hat{\varepsilon}^a_k$ and $\hat{\varepsilon}^b_k$ adopt the form
\bea \label{ferri33} & & \hat{\varepsilon}^a_k\,=\,4 s_1
J_1\,\varepsilon_k\,+\,6\,J\,s_2\mu_1
\nonumber \\
\\
& &
\hat{\varepsilon}^b_k\,=\,4s_2\,J_2\,\varepsilon_k\,+\,6\,J\,s_1\mu_2
\nonumber \eea
We assume $\mu_1$ and $\mu_2$ to be positive
($\mu_1>0,\,\mu_2>0$), then $\hat{\varepsilon}^a_k>0$ and
$\hat{\varepsilon}^b_k>0$ for all values of the wave-vector $k$.
The Bose theory is well defined if square-roots in equations
(\ref{ferri31}) are well defined and $E^{\alpha}_k\geq 0,\quad
E^{\beta}_k\geq 0$. This comes true if \be\label{ferri34} \mu_1
\mu_2\geq1.\ee The $\beta_k$-excitation is gapped ($E^{\beta}_k>0 $)
for all values of parameters $\mu_1$ and $\mu_2$ which satisfy
Eq.(\ref{ferri34}). The $\alpha$-excitation is gapped if $\mu_1
\mu_2>1$, but in the particular case \be \label{ferri35} \mu_1
\mu_2=1\ee $\hat{E}^{\alpha}_0=0$, and near the zero wave vector
\be \label{ferri35b}
\hat{E}^{\alpha}_k\approx \hat{\rho} k^2\ee with
spin-stiffness constant
\be \label{ferri35c} \rho\,=\,\frac {8(s_2^2 J_2\mu_1
\,+\,s_1^2 J_1 \mu_2)\,+\,2s_1s_2J}{(s_1 \mu_2-s_2\mu_1)}\ee In
the particular case Eq.(\ref{ferri35}) $\alpha_k$-boson is the
long-range excitation (magnon) in the system.

\subsection{\bf  $s_1>s_2$ and  $J_1\gg J_2$}

We introduced the parameters $\lambda_1$ and $\lambda_2$ ($\mu_1,
\mu_2$) to enforce the spontaneous magnetization on the sublattices
A and B to be equal to zero. We find out the parameters $\mu_1$ and
$\mu_2$, as functions of temperature, solving the system of two equations $M^A=0$, $M^B=0$ \bea \label{ferri36}& &
s_1\,-\frac 1N \sum\limits_{k\in B_r}\left
[\hat{u}^2_k\,<\alpha^+_k \alpha_k>\,+\,\hat{v}^2_k\,<\beta_k
\beta^+_k>\right]\,=\,0\nonumber
\\
\\
& & -s_2\,+\frac 1N \sum\limits_{k\in B_r}\left
[\hat{u}^2_k\,<\beta^+_k \beta_k>\,+\,\hat{v}^2_k\,<\alpha_k
\alpha^+_k> \right]\,=\,0\nonumber \eea
where the coefficients
$\hat{u}_k$ and $\hat{v}_k$ are given by Eqs(\ref{ferri29}) and the Bose functions (\ref{ferri21}) are defined with dispersions $\hat{E}^{\alpha}_k$ and $\hat{E}^{\beta}_k$. The solutions $\mu_1(T/J)$ and $\mu_2(T/J)$ depend on the parameters $s_1,\,s_2,\,J_1/J$ and $J_2/J$. For $s_1=1.5,\,s_2=1,\,J_1/J=0.47$ and $J_2/J=0.005$ they are depicted in
Fig.2.
\begin{figure}[!ht]
\epsfxsize=9cm 
\epsfbox{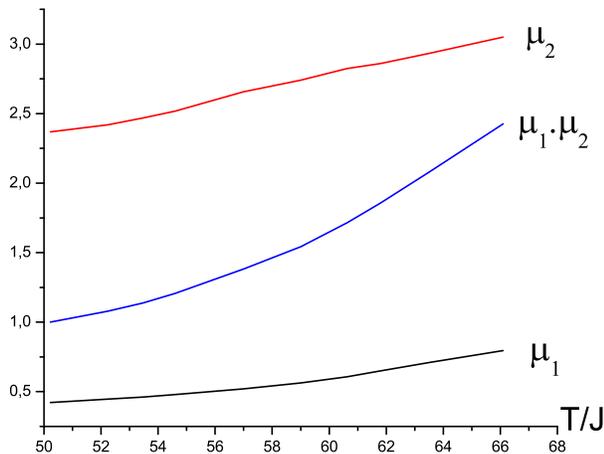} \caption{(color online) The solution of the system of equations (\ref{ferri36}) $\mu_1(T/J)$, $\mu_2(T/J)$ and
$\mu_1(T/J).\mu_1(T/J)$ for parameters $s_1=1.5$,\,$s_2=1$,\,$J_1/J=0.47$ and $J_2/J=0.005$}\label{fig2}
\end{figure}

The upper (red) line is $\mu_2(T/J)$, the bottom (black) line is $\mu_1(T/J)$
and the middle (blue)line is the product $\mu_1(T/J).\mu_2(T/J)$. The
numerical calculations show that
for high enough temperature $\mu_2>1$,\,\, $1>\mu_1>0$ and
$\mu_1.\mu_2>1$. Hence $\alpha_k$ and $\beta_k$ excitations are gapped.
When the temperature decreases $\mu_2$ decreases remaining larger
then one, $\mu_1$ decreases too remaining positive, and  the product
$\mu_1.\mu_2$ decreases remaining larger than one. At
temperature $T_N/J=50.22$ one obtains $\mu_1=0.422$,
$\mu_2=2.368$ and therefor $\mu_1\cdot\mu_1=1$. Hence, at $T_N$
long-range excitation (magnon) emerges in the spectrum which means
that $T_N$ is the Neel temperature.

Below the Neel temperature the spectrum contains magnon
excitations, thereupon $\mu_1\cdot\mu_2=1$. It is convenient to set
\be\label{ferri37}\mu_1=\mu, \quad\quad \mu_2=1/\mu.\ee
\begin{figure}[!h]
\epsfxsize=9cm 
\epsfbox{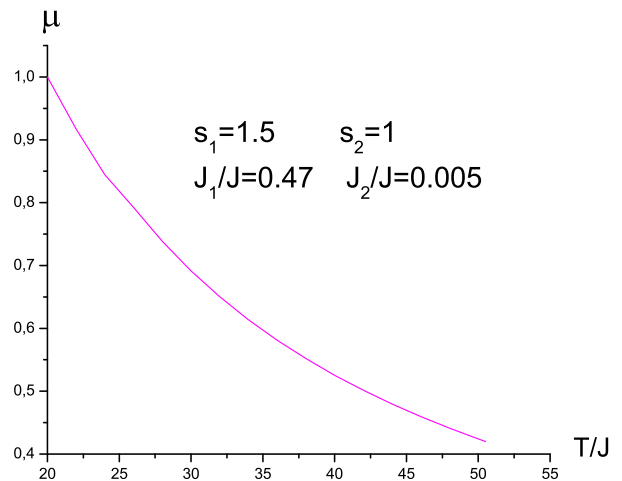} \caption{(color online)Temperature dependence of $\mu$ for $T$ between $T^*$ and $T_N$
and parameters
$s_1=1.5,\,s_2=1,\,J_1/J=0.47$ and $J_2/J=0.005$}\label{fig3}
\end{figure}
In ordered phase magnon excitations are origin of suppression
of the magnetization. Near the zero temperature their contribution is
small and at zero temperature they are close to $s_1$ and $s_2$.  Increasing
the temperature magnon fluctuations suppress the magnetization. For
the chosen parameters they first suppress the sublattice $B$ magnetization
at $T^{*}$ ($M^A(T^{*})>0$). Once suppressed,
the magnetic moment of sublattice $B$ spins can not be restored
increasing the temperature above $T^{*}$. To formulate this
mathematically, we modify the spin-wave theory introducing the
parameter $\mu$ Eq.(\ref{ferri37}). Below $T^{*}$\,\,$\mu=1$,  or
in terms of $\lambda$ parameters $\lambda_1=\lambda_2=0$, which
reproduces the customary spin-wave theory. Increasing the
temperature above $T^*$ the magnetic moment of the sublattice $B$ spins should be zero. This is why we impose the condition
$M^B(T)=0$ if $T>T^{*}$. For temperatures above $T^*$ the
parameter $\mu$ is a solution of this equation. The function $\mu(T/J)$ is depicted in Fig. 3.
Increasing the temperature above $T^*$ $\mu(T/J)$ decreases from $\mu(T^*/J)=1$ to $\mu(T_N/J)=\mu_1(T_N/J)=0.422$.

Next, one utilizes the so
obtained function $\mu(T/J)$ to calculate the sublattice $A$ magnetization
as a function of the temperature. Above
$T^*$ $M^A$ is equal to the magnetization of the system. The
magnetic moments of the sublattice $A$ and $B$ spins, as well as
the magnetization of the system, as a function of the temperature are
depicted in Fig.4 for parameters
$s_1=1.5,\,s_2=1,\,J_1/J=0.47,\,J_2/J=0.005$.
\begin{figure}[!h]
\epsfxsize=9cm 
\epsfbox{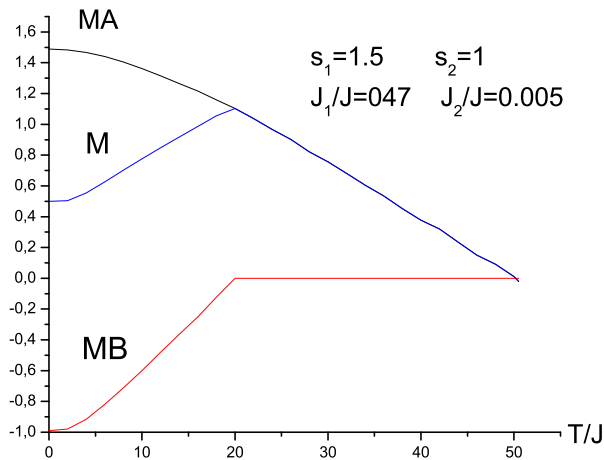} \caption{(color online)Temperature dependence of the ordered
moments: $M$ (blue line)-the magnetization of the system, $M^A$
(black line)-sublattice $A$ magnetization, $M^B$ (red
line)-sublattice $B$ magnetization for parameters
$s_1=1.5,\,s_2=1,\,J_1/J=0.47$ and $J_2/J=0.005$:\, \textbf{modified spin-wave theory} }\label{fig4} \end{figure}

To compare the theoretical results  and the experimental $M(T)$ curves one has, first of all, to interpret adequately the measurements.
The magnetic moments in some materials are close to "spin only" value $2\mu_B S$ and the sublattice spins $s_1$ and $s_2$ can be obtained from the experimental curves. I consider the system $MnCr_2S_{4-x}Se_{x}$. It has been investigated by measurements of the magnetization at $15.3kOe$ as a function of temperature (figure 94 in \cite{HBMM3}). The maximum
in the magnetization versus temperature curve, which is typical of $MnCr_2S_4$ ($x=0$),
increases when $x$ increase, and disappears at $x=0.5$. The Neel temperature decreases
from $74K$ at $x=0$ to $56K$ at $x=2$. The authors' conclusion is that the observed
change of the magnetic properties is attributed to a decrease of the strength
of the negative $Mn^{2+}-Cr^{3+}$ superexchange interaction with increasing $Se$ concentration.

 As follows from the present theory (see Fig.4 middle-blue line) the maximum of the magnetization is at $T^*$. Above $T^*$ the magnetization of the system is equal to the magnetization of sublattice $A$ spins. If we extrapolate this curve below $T^*$ up to zero temperature we will obtain a value close to $2s_1\mu_B$, where $s_1$ is the spin of the sublattice $A$ spin operators. The figure shows that extrapolations give one and just the same result for all values of $x$. One can accept the fact that the $Se$ concentration do not influence over the value of  sublattice $A$ spin and $s_1=1.5$.

Below $T^*$ the magnetization is a sum of sublattice $A$ and $B$ magnetization. Hence, the magnetization at zero temperature is equal to $2(s_1-s_2)\mu_B$. Therefore, one can determine the sublattice $B$ spin $s_2$. The important conclusion is that the effective sublattice $B$ spin $s_2$ decreases with increasing Se concentration. The dimensionless magnetization (in Bohr magnetons) per lattice site $\sigma=2(M^A+M^B)$ is calculated for $s_1=1.5,\,J_1/J=0.47,\,J_2/J=0.005$ and $s_2=1;\, 0.7;\, 0.4$. The curves are depicted in Fig.5.
\begin{figure}[!h]
\epsfxsize=9cm 
\epsfbox{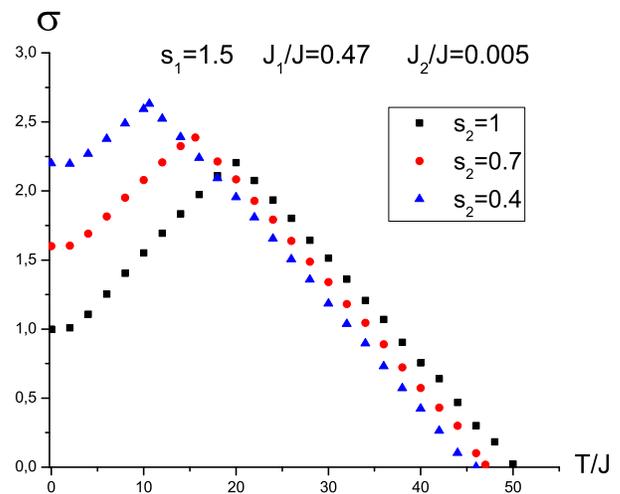} \caption{(color online)The magnetization, in Bohr magnetons, per lattice site $\sigma=2(M^A+M^B)$ for $s_1=1.5,\,J_1/J=0.47,\,J_2/J=0.005$ and $s_2=1;\, 0.7;\, 0.4$}\label{fig5}
\end{figure}
The figure shows that the present calculations capture the essential  features of the system; increasing $Se$ concentration (decreasing $s_2$) leads to decreasing of  Neel temperature, $T^*$ temperature decreases too, and the maximum of the magnetization $\sigma(T^*)$ increases.

\subsection{\bf $s_1>s_2$ and  $J_2\gg J_1$}

Next I consider the case when the sublattice $A$ spin $s_1$ is larger then the sublattice $B$ spin $s_2$ but the intra-sublattice exchange constant $J_1$ is much smaller than the intra-sublattice $B$ exchange constant $J_2$. Increasing the temperature, the magnon fluctuations suppress the magnetic order and sublattice $A$ magnetization decreases faster than sublattice $B$ magnetization. There is a temperature $T_c$ at which the magnetization of the system is zero $M(T_c)=0$ (the compensation point). Increasing the temperature above $T_c$ the sublattice $A$ magnetization becomes equal to zero at $T^*$. Above this temperature the sublattice $A$ magnetization should be kept equal to zero and one utilizes the modified spin-wave theory to calculate the magnetization of the system which is equal to sublattice $B$ magnetization. The magnetization curves $M^A(T/J),\,M^B(T/J)$ and $M(T/J)$ are depicted in Fig. 6 for parameters $s_1=18,\,s_2=3,\,j_1=J_1/J=0.005$ and $j_2=J_2/J=1$.
\begin{figure}[!h]
\epsfxsize=9cm 
\epsfbox{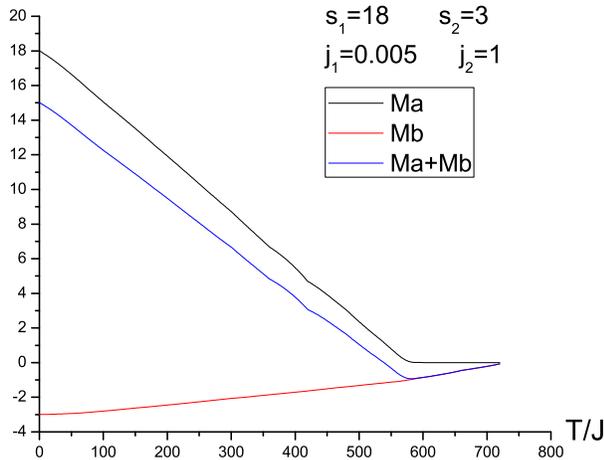} \caption{(color online)\,The magnetization curves:\,$M^A(T/J)$\,-\,upper (black) line,\,$M^B(T/J)$\,-\,bottom (red) line, $M(T/J)$\,-\,middle (blue) line for parameters $s_1=18,\,s_2=3,\,j_1=J_1/J=0.005$ and $j_2=J_2/J=1$: \textbf{modified spin-wave theory}}\label{fig6}
\end{figure}

To compare the theoretical results with experimental ones I address the magnetization-temperature
curves of rare earth iron garnets.When a rare earth ion, such as $Gd^{3+}$ through $Yb^{3+}$, is present
instead of a diamagnetic ion, such as $Y^{3+}$ in Yttrium Iron Garnet, it is found that the
magnetization as a function of temperature shows compensation points, i.e.,
temperatures at which the spontaneous magnetization is zero (Fig. 5 \cite{Wolf}). I use the "spin only" value for the magnetization, and the sublattices' spins are consider as an effective parameters to be determined from the experimental curves. Usually the magnetization, in Bohr magnetons, per lattice site $\sigma=2|M^A+M^B|$ are depicted in figures. $T^*$ is the temperature at which the sublattice $A$ magnetization becomes equal to zero (black line Fig.6), and hence the magnetization is minimal (blue line Fig. 6), in the case of the $\sigma(T)$ curves $T^*$ is the temperature, above the compensation point, at which $\sigma(T^*)$ is maximal. The magnetization above $T^*$ is equal to the sublattice $B$ magnetization. To assess the value of the sublattice $B$ spin $s_2$, one has to extrapolate the curve to zero temperature. Then the spin $s_1$ can be obtained from the zero temperature of the magnetization $2(s_1-s_2)\mu_B$. The experimental curves (Fig. 5 \cite{Wolf}) show that the value of $\sigma(T^*)$ is different for different rare earth ions, and hence the effective spin $s_2$ is different for different rare earths. It is smaller for gadolinium, increases for terbium, dysprosium, holmium, and is higher for ytterbium. At the same time the temperature $T^*$ decreases, which means that sublattice $B$ intra-exchange constant $J_2$ decreases. The magnetization   $\sigma=2|M^A+M^B|$ for two choices of the parameters are depicted in Fig.7: black squares-$s_1=18,\,s_2=3,\,j_1=J_1/J=0.005,\,j_2=J_2/J=1$ and red circles-$s_1=18,\,s_2=6,\,j_1=J_1/J=-0.05,\,j_2=J_2/J=0.15$.
\begin{figure}[!h]
\epsfxsize=9cm 
\epsfbox{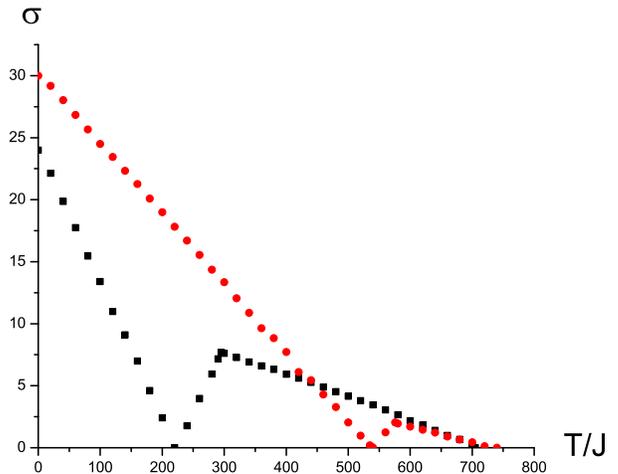} \caption{(color online)The magnetization $\sigma=2|M^A+M^B|$ as a function of $T/J$: black squares\,-\,$ s_1=18,\,s_2=3,\,\,J_1/J=0.005,\,J_2/J=1$, red circles\,-\,$s_1=18,\,s_2=6,\,\,J_1/J=-0.05,\,J_2/J=0.15$}\label{fig7}
\end{figure}
The negative sign of $J_1$ in the second group of parameters is chosen to reproduce the shape of the experimental curves. One can do better fitting, with positive constants $J_2$ for all rare earths, accounting for magnon scattering processes.

\subsection{\bf $s_1=s_2$  and  $J_1\gg J_2$}

Finally I consider a theory with parameters $s_1=s_2=s$ and $J_1\gg J_2$. It is more correctly to thought of this case as a generalization of the antiferromagnetism. The system has two magnons with dispersions $E_k^{\alpha}$ and $E_k^{\beta}$. Near the zero wave vector the dispersions' asymptotic is as in the antiferromagnetic case
\be \label{ferri37} E^{\alpha}_k\approx v_s |{\bf k}|, \qquad E^{\beta}_k\approx v_s |{\bf k}|\ee with spin-wave velocity
\be \label{ferri38} v_s=2s\sqrt{12J_1\,+\,12J_2\,+\,3J}.\ee As a result, at zero temperature and near the zero temperature one observes a compensation of sublattice $A$ and $B$ magnetization  and magnetization of the system is zero. The difference of the dispersions is
\be \label{ferri39} E^{\alpha}_k\,-\,E^{\beta}_k\,=\,4s(J_1-J_2)\varepsilon_k\approx (J_1-J_2){\bf k}^2, \ee
therefore increasing the temperature one obtains non-compensation of sublattice $A$ and $B$ magnetization. The magnon fluctuations suppress first the sublattice $B$ magnetization at temperature $T^*$. Above this temperature only sublattice $A$ contributes the magnetization. The magnetization curves $M^A(T/J),\,M^B(T/J)$ and $M(T/J)$ are depicted in Fig. 8 for parameters $s_1=1,\,\,s_2=1,\,\,J_1/J=0.47$ and $J_2/J=0.005$.
\begin{figure}[!th]
\epsfxsize=9cm 
\epsfbox{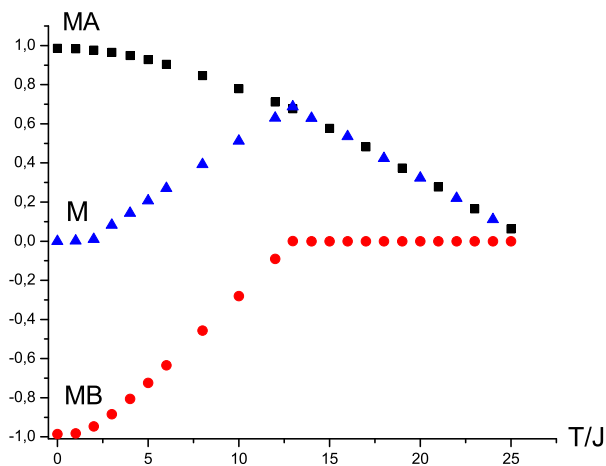} \caption{(color online)\,The magnetization curves:\,$M^A(T/J)$\,-\,black squares,\,$M^B(T/J)$\,-\,red circles, $M(T/J)$\,-\,blue triangles for parameters $s_1=1,\,s_2=1,\,J_1/J=0.47$ and $J_2/J=0.005$: \textbf{modified spin-wave theory} }\label{fig8}
\end{figure}
The magnetization-temperature curve (blue triangles) reproduces the Neel's anomaly.

\section{Summary}

In summary, I have calculated the magnetization as a function of temperature for two-sublattice ferrimagnet. Anomalous temperature dependence of the magnetization, predicted by Neel, is reproduced, and the shape of the theoretical curves satisfactorily coincides with the experimental ones.
The most important difference between Neel's theory and the present modified spin-wave theory is that Neel's calculations predict a temperature $T_N$ at which both the sublattice $A$ and $B$ magnetizations become equal to zero. The modified spin wave theory predicts to phases: at low temperatures $(0,T^*)$ the magnetic orders of the two sublattices contribute to the
magnetization of the system, while at  high temperatures  $(T^*,T_N)$ only one of the sublattices, that with stronger intra-sublattice exchange, has non-zero spontaneous magnetization. It is important to stress that despite the fact that only one of the spins contribute the magnetic order at high temperatures $T^*<T<T_N$ this is not a ferromagnetic phase. This is a ferrimagnetic phase because the magnon of the system is a mixture of the transversal fluctuations of the two spins.

Two ferromagnetic phases where theoretically predicted, very recently,  in spin-Fermion systems, which obtain their magnetic properties
from a system of localized magnetic moments being coupled to
conducting electrons \cite{Karchev1}. At the characteristic
temperature $T^{*}$, the magnetization of itinerant electrons becomes
zero, and high temperature ferromagnetic phase ($T^{*}<T<T_C$) is a
phase where only localized electrons give contribution to the
magnetization of the system. An anomalous increasing of
magnetization below $T^{*}$ is obtained in good agrement with experimental measurements of the ferromagnetic phase of $UGe_2$\cite{2fmp6}.
The results of the present paper and the previous one \cite{Karchev1} suggest that $T^*$ transition from  a magnetic phase to another magnetic phase is a generic feature of the two spin systems. The additional phase transition demonstrates itself through the anomalous temperature variation of the spontaneous magnetization. Another way is possible, to experimentally identify the transition, if $T^*$ is a temperature at which the itinerant electrons in the system start to form magnetic moment. We know from itinerant ferromagnetism that the transition is observed through the change in the temperature dependence of resistivity. The most prominent example is the spinel $Fe_3O_4$. It has been extensively investigated and most striking feature is the Verwey transition \cite{Verwey}. At relatively low temperature $T_t=100-120 K$ the magnetization abruptly changes the slope, and the conductivity has anomalous behavior. Many decades of research on $Fe_3O_4$ have led to the view \cite{Belov2} that the spinel is a two-sublattice system with three spins. The $Fe_A^{3+}$ and $Fe_B^{3+}$ ions are associated with localized spins on sublattices $A$ and $B$, while $Fe_B^{2+}$ ions are associated with itinerant electrons on sublattice $B$. With this in mind one can interpret $T_t$ as a temperature at which the magnetic moment of itinerant electrons sets in.

The present theory of ferrimagnetism permits to formulate phenomenological rules which to enable the analysis  of experimental results for more complicate systems without explicit calculations. For example, the three-spin systems have, most generally, three characteristic temperatures $T^*_1<T^*_2<T_N$.
The Neel temperature $T_N$ is the temperature at which the magnetic order sets in. When $T^*_2<T<T_N$ only one of the spins has non zero spontaneous magnetization and the magnetization of the system equals the magnetic order of that spin. When $T^*_1<T<T^*_2$ two of the spins contribute the magnetization of the system. If $M(T^*_2)$ is local maximum the two spins are with anti-parallel order. If the
magnetization shows an anomalous enhancement below $T^{*}_2$ the two spins are parallel. Below $T^*_1$ all three spins have non zero spontaneous magnetization, and the magnetization changes the slope at this temperature.

As an example I consider the $CeCrSb_3$ compound \cite{Jackson1}. The magnetic behavior is determined by $Cr$ ions and $Ce-4f$ electrons, which are sat in different sublattices. The measurements of the magnetization and electric resistivity on single crystal of $CeCrSb_3$ show that there are three characteristic temperatures, Neel temperature $T_N=115K$, $T^*_2=108K$ and $T^*_1=18K$. The closeness of $T_N$ and $T^*_2$ leads to misinterpretation and even to identification of these temperatures, but evident increase of magnetization when $T^*_2<T<T_N$ and subsequent decrease below $T^*_2$ proves  that there are three characteristic temperatures. Below $T^*_1$ the magnetization increases again. The temperature variation of the resistivity shows anomalous behavior at $T_N$ and $T^*_1$. The experimental results suggest that part of the itinerant $Cr$ electrons start to form magnetic order at Neel temperature $T_N$ and the rest one at $T^*_1$, while the localized $Ce-4f$ electrons do this at $T^*_2$.
At high temperature  $(T^*_2,T_N)$ the magnetic order of some of the $Cr$ electrons contribute the magnetization of the system. Below $T^*_2$ the the magnetic moment of localized  $Ce$ electrons sets in. The low temperature phase $(T^*_1,T^*_2)$ is a phase where all  $Cr$ and $Ce$ electrons contribute the magnetization of the system.

Another example is the manganese vanadium oxide spinel $MnV_2O_4$ \cite{vanadium1}. On one of the sublattices sites is $Mn^{+2}$ in a $3d^5$ high spin configuration $s=5/2$ with quenched orbital angular momentum. On the sites of the other sublattices are the vanadium electrons. More precise measurements of the magnetization as a function of temperature \cite{vanadium2} show that the set in of magnetic order is due to $Mn$ localized electrons at Neel temperature $T_N=56.5K$. Because of the splitting of the vanadium three bands some of electrons start to form magnetic moment at $T^*_2=48K$. An evidence for this is the abrupt decrease of magnetization below $T^*_2$, which also indicates that the magnetic order of vanadium electrons is anti-parallel with the order of $Mn$ electrons. Below $T^*_1=27K$ the magnetization decreases faster which means that all vanadium electrons are involved in the magnetization of the system.
\begin{center}
\textbf{Acknowledgements}
\end{center}
The author wants to thank Professor Suzuki for valuable discussions.



\begin{thebibliography}{99}
%
\bibitem[*]{byline} Electronic address: naoum@phys.uni-sofia.bg
\bibitem{Neel} L. Neel,  Ann. Phys., Paris, {\bf 3}, 137 (1948).
\bibitem{Wolf} W. P. Wolf, Rep. Prog. Phys., {\bf 24}, 212 (1961).
\bibitem{HBMM1} K. H. J. Buschow, in \emph{Handbook of Magnetic Materials},
Volume {\bf 1}, 297, Edited by E.P Wohlfarth, (North-Holland Publishing Company, 1980).
\bibitem{HBMM2} M. A. Gilleo, in \emph{Handbook of Magnetic Materials},
Volume {\bf 2}, 1, Edited by E.P Wohlfarth, (North-Holland Publishing Company, 1980).
\bibitem{HBMM3} P. P. Van Stapele, in \emph{Handbook of Magnetic Materials},
Volume {\bf 3}, 603, Edited by E.P Wohlfarth, (North-Holland Publishing Company, 1982).
\bibitem{Belov} K. P. Belov, Uspekhi Fizicheskikh Nauk, {\bf 166}, 669 (1996) [Physics-Uspekhi, {\bf 39}, 623 (1996)].
\bibitem{swKaplan} H. Kaplan, Phys.Rev. {\bf 88}, 121 (1952).
\bibitem{itinerant} P. P. Craig, W. I. Goldburg, T. A. Kitchens, and J.
I. Budnick, Phys. Rev. Lett., {\bf 19}, 1334 (1967).
\bibitem{Takahashi1} M.Takahashi, Prog. Theor. Physics Supplement {\bf 87}, 233 (1986).
\bibitem{Takahashi2} M.Takahashi, Phys. Rev. Lett. {\bf 58}, 168 (1987).
\bibitem{Karchev1} N. Karchev, Phys. Rev. {\bf B 77}, (2008)[arXiv:cond/mat-0709.1759 (2007)].
\bibitem{2fmp6} C. Pfleiderer and A. D. Huxley, Phys. Rev. Lett., {\bf 89}, 147005 (2002).
\bibitem{Verwey} E. J. W. Verwey, Nature (London) {\bf 144}, 327 (1939).
\bibitem{Belov2} K. P. Belov, Uspekhi Fizicheskikh Nauk, {\bf 163}, 53 (1993) [Physics-Uspekhi, {\bf 37}, 563 (1994)].
\bibitem{Jackson1} D. D. Jackson, S. K. McCall, A. B. Karki, and D. P. Young, Phys. Rev. {\bf B 76}, 064408 (2007).
\bibitem{vanadium1} K. Adachi, T. Suzuki, K. Kato, K. Osaka, M. Takata, and T. Katsufuji,  Phys. Rev. Lett., {\bf 95}, 197202 (2005).
\bibitem{vanadium2} V. O. Carlea, R. Jin, D. Mandrus, B. Roessli, Q. Huang, M. Miller, A. J. Schultz, and S. E. Nagler, arXiv:0711.1844 (2007).



\end{thebibliography}
\end{document}